\documentclass[10pt,aps,prl,floatfix,twocolumn,superscriptaddress,nofootinbib]{revtex4-1}

\usepackage[labelfont={small},subrefformat=parens,caption=false]{subfig}
\usepackage[dvipsnames]{xcolor}
\usepackage{amsfonts,amsmath,amssymb,bm}
\usepackage{graphicx}
\usepackage[utf8]{inputenc}
\usepackage{xspace}
\usepackage{isotope}
\usepackage{xparse}
\usepackage{physics}

\usepackage[pdfpagelabels, pdfencoding=auto, psdextra]{hyperref}
\hypersetup{%
 pdfsubject=Paper,
 pdfkeywords={nuclear physics} {Bayesian analysis} {chiral EFT} {nuclear matter} {machine Leaning},
 unicode = true,
 breaklinks = true,
 colorlinks = true,
 linkcolor = blue,
 menucolor = blue,
 citecolor = blue,
 urlcolor = blue
}

\makeatletter
\let\origcitation\citation
\AtEndDocument{\def\mycites{}%
	\def\citation#1{\g@addto@macro\mycites{#1^^J}\origcitation{#1}}}
\AtVeryEndDocument{\newwrite\citeout\immediate\openout\citeout=\jobname.cit
	\immediate\write\citeout{\mycites}\immediate\closeout\citeout}
\makeatother

\definecolor{bobcatgreen}{rgb}{0.3,0.6,0.1}
\definecolor{scarlet}{rgb}{0.73,0,0}

\graphicspath{{./figures/}}

\usepackage{silence}
\makeatletter
\newcommand\newsubcommand[3]{\newcommand#1{#2\sc@sub{#3}}}
\def\sc@sub#1{\def\sc@thesub{#1}\@ifnextchar_{\sc@mergesubs}{_{\sc@thesub}}}
\def\sc@mergesubs_#1{_{\sc@thesub#1}}

\newcommand\newsupcommand[3]{\newcommand#1{#2\sc@sup{#3}}}
\def\sc@sup#1{\def\sc@thesup{#1}\@ifnextchar^{\sc@mergesups}{^{\sc@thesup}}}
\def\sc@mergesups^#1{^{\sc@thesup#1}}
\makeatother

\DeclareMathAlphabet{\mathbcal}{OMS}{cmsy}{b}{n}

\newcommand{\etal}{\textit{et~al.}\xspace}
\newcommand{\Msol}{\,\text{M}_{\odot}\xspace}

\newcommand{\fmi}{\, \text{fm}^{-1}}
\newcommand{\fmiq}{\, \text{fm}^{-3}}
\newcommand{\keV}{\, \text{keV}}
\newcommand{\MeV}{\, \text{MeV}}

\newcommand{\NNLO}{\ensuremath{{\rm N}{}^2{\rm LO}}\xspace}
\newcommand{\NNNLO}{\ensuremath{{\rm N}{}^3{\rm LO}}\xspace}

\newcommand{\ordervec}{\vec}

\newcommand{\inputvec}{\mathbf}

\newsubcommand{\ckvec}{\ordervec{c}}{k}

\newsubcommand{\bkvec}{\ordervec{b}}{k}

\newsubcommand{\ckvecset}{\ordervec{\inputvec{c}}}{k}

\newsubcommand{\ckvecapprox}{\mathbf{c}'}{k}
\newsubcommand{\ckvecapproxset}{\mathbf{C}'}{k}

\newsubcommand{\bkvecapprox}{\mathbf{b}'}{k}
\newsubcommand{\bkvecset}{\mathbf{B}}{k}
\newsubcommand{\bkvecapproxset}{\mathbf{B}'}{k}

\newcommand{\genobs}{y}

\newsubcommand{\genobsvec}{\ordervec{\genobs}}{k}
\newsubcommand{\genobsvecset}{\ordervec{\inputvec{\genobs}}}{k}

\newsubcommand{\akvec}{\mathbf{a}}{k}

\newsubcommand{\akvecapprox}{\mathbf{a}'}{k}
\newsubcommand{\akvecset}{\mathbf{A}}{k}
\newsubcommand{\akvecapproxset}{\mathbf{A}'}{k}

{}  

\DeclareMathOperator{\pr}{pr} 
\newcommand{\given}{\,|\,}  

\newcommand{\chiEFT}{$\chi$EFT}

\newcommand{\kf}{\ensuremath{k_{\scriptscriptstyle\textrm{F}}}}

\def\diffd{\mathrm{d}}  

\DeclareDocumentCommand\differential{ o g d() }{ 
    \IfNoValueTF{#2}{
        \IfNoValueTF{#3}
            {\diffd\IfNoValueTF{#1}{}{^{#1}}}
            {\mathinner{\diffd\IfNoValueTF{#1}{}{^{#1}}\argopen(#3\argclose)}}
        }
        {\mathinner{\diffd\IfNoValueTF{#1}{}{^{#1}}#2} \IfNoValueTF{#3}{}{(#3)}}
    }
\DeclareDocumentCommand\dd{}{\differential} 

\newcommand{\pathd}{\mathcal{D}}  

\DeclareDocumentCommand\pathdifferential{ o g d() }{ 
    \IfNoValueTF{#2}{
        \IfNoValueTF{#3}
            {\pathd\IfNoValueTF{#1}{}{^{#1}}}
            {\mathinner{\pathd\IfNoValueTF{#1}{}{^{#1}}\argopen(#3\argclose)}}
        }
        {\mathinner{\pathd\IfNoValueTF{#1}{}{^{#1}}#2} \IfNoValueTF{#3}{}{(#3)}}
    }

\newcommand{\phantomsublabel}[3]{%
\unitlength=1in%
\put(#1,#2){%
 \subfloat[]{%
 \label{#3}%
 }}%
}

\newcommand{\prlsection}[1]{\emph{#1.}---}
\newcommand{\verifyvalue}[1]{#1}
\newcommand{\eg}{\textit{e.g.}\xspace}
\newcommand{\ie}{\textit{i.e.}\xspace}

\begin{document}


\title{How Well Do We Know the Neutron-Matter Equation of State at the Densities
Inside Neutron Stars? A Bayesian Approach with Correlated Uncertainties}

\author{C. Drischler}
\email{cdrischler@berkeley.edu}
\affiliation{Department of Physics, University of California, Berkeley, California 94720, USA}
\affiliation{Nuclear Science Division, Lawrence Berkeley National Laboratory,
Berkeley, California 94720, USA}

\author{R.~J. Furnstahl}
\email{furnstahl.1@osu.edu}
\affiliation{Department of Physics, The Ohio State University, Columbus, Ohio
43210, USA}

\author{J.~A. Melendez}
\email{melendez.27@osu.edu}
\affiliation{Department of Physics, The Ohio State University, Columbus, Ohio
43210, USA}

\author{D.~R. Phillips}
\email{phillid1@ohio.edu}
\affiliation{Department of Physics and Astronomy and Institute of Nuclear and
Particle Physics, Ohio University, Athens, Ohio 45701, USA}

\date{\today}

\begin{abstract}

We introduce a new framework for quantifying correlated uncertainties of the
infinite-matter equation of state derived from chiral effective field
theory ($\chi$EFT). Bayesian machine learning via Gaussian processes with
physics-based hyperparameters allows us to efficiently quantify and propagate
theoretical uncertainties of the equation of state, such as $\chi$EFT truncation errors, to
derived quantities. We apply this framework to state-of-the-art many-body
perturbation theory calculations with nucleon-nucleon and
three-nucleon interactions up to fourth order in the $\chi$EFT expansion. This
produces the first statistically robust uncertainty estimates for key
quantities of neutron stars. We give results up to twice nuclear saturation
density for the energy per particle, pressure, and speed of sound of neutron
matter, as well as for the nuclear symmetry energy and its derivative. At
nuclear saturation density the predicted symmetry energy and its slope are
consistent with experimental constraints.

\end{abstract}

\maketitle

\prlsection{Introduction}%
How well do we know the neutron-matter equation of state (EOS) at the densities
inside neutron stars? This is a key question for nuclear (astro)physics in the
era of multimessenger astronomy. To answer this question from nuclear
theory requires a systematic understanding of strongly interacting, neutron-rich
matter at densities several times the typical density in heavy nuclei, \ie, well
beyond the nuclear saturation density $n_0 \approx 0.16 \fmiq$ ($\rho_0 \approx 2.7
\times 10^{14} \, \text{g\,cm}^{-3})$. The dominant microscopic approach to
describing nuclear forces at low energies is chiral effective field theory
(\chiEFT) with nucleon and pion degrees of freedom~\cite{Epel09RMP, Mach11PR,
Hammer:2019poc, Tews:2020hgp}. It has made great progress in predicting the EOS of infinite
(nuclear) matter and the structure of neutron stars at densities $\lesssim
n_0$~\cite{Tews13N3LO, Hebe13ApJ,Baar13CCinf,Drischler:2013iza, Hagen:2013yba,Carbone:2013rca,Coraggio:2014nva, Well14nmtherm, Rogg14QMC, Holt:2016pjb,Drischler:2016djf,Ekstrom:2017koy, Drischler:2017wtt, Lonardoni:2019ypg, Piarulli:2019pfq} (see also
Refs.~\cite{Hebe15ARNPS, Drischler:2019xuo, Sammarruca:2019ncy} for recent reviews). But the
truncation errors inherent in \chiEFT\ grow dramatically with
density~\cite{Leonhardt:2019fua, Lonardoni:2019ypg, Tews19gw, Tews18gw, Tews18sos}. Existing
predictions only provide rough estimates for them and do not account for
correlations within or between observables.

In this Letter we use a novel Bayesian approach to quantify the truncation
errors in \chiEFT\ predictions for pure neutron matter (PNM) at zero
temperature~\cite{Melendez:2019izc,Drischler:2020preparation}. The EOS is
obtained from state-of-the-art many-body perturbation theory (MBPT) calculations
with nucleon-nucleon (NN) and three-nucleon (3N) interactions up to
fourth order in the \chiEFT\ expansion (\ie, next-to-next-to-next-to-leading
order, \NNNLO)~\cite{Drischler:2017wtt}. Our algorithm accounts for correlations in EOS
truncation errors---both across densities and between
observables---enabling us to obtain reliable uncertainties for physical
properties derived from the EOS, \eg, the pressure and the speed of sound. This
significant advance in uncertainty quantification (UQ) is timely given the need
for statistically robust comparisons~\cite{PhysRevA.83.040001} between nuclear theory and recent
observational constraints~\cite{LIGO17MultMess, LIGO18NSradii, LIGO18update, 
Bogd19NICER2, Riley19NICER, Mill19NICER, Raai19NICER,Essick:2020flb}.

\chiEFT\ is a systematic expansion in powers of a typical momentum scale, $p$,
over the EFT breakdown scale, $\Lambda_{b}$. For infinite matter, $p$ is of
order the nucleon Fermi momentum $\kf$. 
Provided $\kf < \Lambda_b$, \chiEFT\ calculations of strongly interacting matter can be improved to any desired accuracy. 
In practice there is always a discrepancy between the
\chiEFT\ result and reality, because observables are calculated at a
finite order in the expansion, leaving a residual error that must be
quantified~\cite{Furnstahl:2014xsa, Furnstahl:2015rha, Wesolowski:2015fqa}.

Melendez~\etal~\cite{Melendez:2019izc} developed a Bayesian model for EFT
truncation errors that accounts for uncertainties that vary smoothly with 
independent variable(s)---in this case $\kf$ or the nucleon density
$n$. A machine-learning algorithm is trained on the computed orders in the
\chiEFT\ expansion, from which it learns the magnitude of the truncation error and
its correlations in density.
In a companion publication~\cite{Drischler:2020preparation}, we apply this new
approach to infinite matter. This provides the first
estimates of in-medium EFT breakdown scales and nuclear saturation properties
with correlated EFT truncation errors. We also uncover a strong correlation
between PNM and symmetric nuclear matter (SNM) for the \chiEFT\ Hamiltonians of
interest. This is crucial to the UQ of the nuclear symmetry energy we present here.

This Letter focuses on PNM and, together with its companion
paper~\cite{Drischler:2020preparation}, sets a new standard for UQ in
infinite-matter calculations based on \chiEFT. (See Section~4.2 in Ref.~\cite{Bedaque:2020pct} for an overview of other recent applications of machine learning and Bayesian methods in low-energy nuclear theory.) We first review definitions relevant to our study:
the energy per particle, pressure, and speed of sound, along with the symmetry
energy and its slope. Next we explain how machine-learning algorithms can
estimate statistically robust, correlated uncertainties for these observables. We then provide our results and show that,
for the symmetry energy and its slope at saturation density, they are in
accord with experimental and theoretical constraints.
The annotated Jupyter notebooks we used for the UQ of infinite-matter
observables and their derivatives are publicly available~\cite{BUQEYEgithub}.

\prlsection{Equation of State}%
We consider the standard (quadratic) expansion
of the infinite-matter EOS as a function of the total density $n = n_n + n_p$
and isospin asymmetry $\beta = (n_n-n_p)/n$,
\begin{equation} \label{eq:EA}
 \frac{E}{A}(n, \, \beta) \approx \frac{E}{A}(n,
 \, \beta = 0) + \beta^2 \, S_2(n) \,,
\end{equation}
about SNM ($\beta=0$); with the neutron (proton) density given by $n_n$ ($n_p$).
Microscopic asymmetric matter calculations based on chiral NN and 3N
interactions at $n \lesssim n_0$ have shown that this  expansion
works reasonably well~\cite{Drischler:2013iza,Drischler:2015eba} (cf.
Refs.~\cite{Kaise15quartic,Well16DivAsym}). The density-dependent symmetry
energy $S_2(n)$ is then given by the difference between the energy per particle
in PNM ($\frac{E}{N}$) and SNM ($\frac{E}{A}$),
\begin{equation} \label{eq:S2}
	S_2(n) \approx \frac{E}{N}(n) - \frac{E}{A}(n) \equiv S_v + \frac{L}{3} \left(
	\frac{n-n_0}{n_0} \right)+ \ldots \, .
\end{equation}

We focus our analysis on four key quantities for PNM and neutron stars. The
first two are $S_2(n)$ and its (rescaled) density-dependent derivative
$L(n)\equiv 3n\,\dv{n} S_2(n)$. When evaluated at $n_0$ these become,
respectively, $S_v \equiv S_2(n_0)$ and $L\equiv L(n_0)$. The other two
quantities are the pressure
\begin{equation} \label{eq:P}
	P(n) = n^2 \dv{n} \frac{E}{N}(n) \, ,
\end{equation}
and the speed of sound squared,
\begin{equation} \label{eq:cs2}
	c_s^2(n) = \frac{\partial P(n)}{\partial \varepsilon(n)} = \frac{\partial
	P(n)}{\partial n} \left[ \left( 1 + n \frac{\partial}{\partial n} \right)
	\frac{E}{N}(n) + m_n \right]^{-1} .
\end{equation}
Note that the energy density $\varepsilon(n) = n \left[ \frac{E}{N}(n) + m_n
\right]$ includes the neutron rest mass energy $m_n$ (with $c^2 = 1$).

The central many-body inputs of our analysis are $\frac{E}{N}(n)$ and
$\frac{E}{A}(n)$ as obtained in MBPT\@. We extend the neutron-matter
calculations in Ref.~\cite{Drischler:2017wtt} to $2n_0$ and use the results reported
in Refs.~\cite{Drischler:2017wtt, Leonhardt:2019fua} for SNM\@. The high-order MBPT
calculations are driven by the novel Monte Carlo framework introduced by
Drischler~\etal~\cite{Drischler:2017wtt}. It uses automatic code generation to
efficiently evaluate arbitrary interaction and many-body diagrams, facilitating
calculations with controlled many-body uncertainties (for details see Ref.~\cite{Drischler:2017wtt}).

Reference~\cite{Drischler:2017wtt} also constructed a family of
order-by-order chiral NN and 3N potentials up to \NNNLO. The NN potentials by
Entem, Machleidt, and Nosyk~\cite{Entem:2017gor} with momentum cutoffs $\Lambda
= 450$ and $500\MeV$ were combined with 3N interactions at the same order and cutoff. The
two 3N low-energy couplings $c_{D}$ and $c_{E}$ were fit to the triton and the
empirical saturation point of SNM. These intermediate- and short-range 3N interactions, respectively, do not
contribute to $\frac{E}{N}(n)$ with nonlocal regulators~\cite{Hebeler:2009iv}.
There is consequently only one neutron-matter EOS determined for each cutoff and
\chiEFT\ order~\cite{Drischler:2017wtt}. Our results for a given cutoff do not differ significantly for
the different 3N fits.
We restrict the discussion here to the $\Lambda =
\verifyvalue{500}\MeV$ potentials of Ref.~\cite{Drischler:2017wtt} with $c_D =
\verifyvalue{-1.75}$ $(\verifyvalue{-3.00})$ and $c_E = \verifyvalue{-0.64}$
$(\verifyvalue{-2.22})$ at \NNLO (\NNNLO) and refer to the Supplemental\
Material~\cite{SuppMat} for results with the other cutoff. More details on these Hamiltonians
can be found in Refs.~\cite{Drischler:2017wtt, Drischler:2020preparation}.

\prlsection{Uncertainty Quantification}%
Our truncation-error model relies on Gaussian processes (GPs), a
machine-learning algorithm, to uncover the size and smoothness properties of the
EFT uncertainty~\cite{rasmussen2006gaussian}. We train physically motivated GPs
from our UQ framework to the order-by-order predictions of $\frac{E}{N}(n)$ and
$\frac{E}{A}(n)$, leading to smooth regression curves. Training refers here to
both estimating the GP hyperparameters (\eg, $\Lambda_{b}$ and the GP
correlation length) and finding the regression curve. Note that this
requires choices for the functional form of the EFT expansion  parameter and a
reference scale for each observable, as discussed in
Refs.~\cite{Melendez:2019izc, Drischler:2020preparation}. The resulting curves
also include an interpolation uncertainty that accounts for many-body
uncertainties in the training data. Reference~\cite{Drischler:2017wtt} showed that
the residual many-body uncertainty is much smaller than the \chiEFT\ truncation
error for the interactions considered here. Nevertheless, to be conservative, we
set this additional interpolation uncertainty to \verifyvalue{0.1\%} of the total
energy per particle (but $\geqslant \verifyvalue{20}\keV$). The
results are insensitive to that choice.

An important byproduct of finding the optimal regression curves is a Gaussian
posterior for the truncation error that includes correlations in density.
Combining the respective regression curve with the interpolation and EFT
truncation uncertainties produces a GP for each EOS from the to-all-orders
EFT\@. Furthermore, GPs are closed under differentiation. It is then
straightforward to compute a joint distribution that includes correlations
between the EOS and its derivatives~\cite{Rasmussen:2003gphmc, Solak:2003dgpds,
Eriksson:2018scaling,Chilenski_2015_gptools}.

But assessing the uncertainty in $S_2(n)$ requires an additional step. We have
found that $\frac{E}{N}(n)$ and $\frac{E}{A}(n)$ converge in a similar
fashion~\cite{Drischler:2020preparation}; given an EFT correction of
$\frac{E}{N}(n)$, it is likely that the correction to $\frac{E}{A}(n)$ will have
the same sign. This additional correlation \emph{between} observables implies
that the truncation error in $S_2(n)$ is less than the in-quadrature sum of
errors from PNM and SNM\@. Our truncation framework naturally extends to this
case via \emph{multitask} machine learning (for details see
Ref.~\cite{Drischler:2020preparation}; also~\cite{alvarez2012kernels,
    Melkumyan:2011multikernel, Caruana:1997MultitaskLearning,
    Zhang:2018MultitaskLearning}). The correlation found with multitask
GPs precisely matches the empirical correlation.

Our novel framework thus permits the efficient evaluation of arbitrary derivatives
and the full propagation of uncertainties within and between observables. Each
type of correlation is essential for full UQ in infinite matter: without
correlations in density, derivatives of the EOS would have grossly
exaggerated uncertainties; without correlations between observables and their
derivatives, UQ for $c_s^2(n)$ would not be reliable; without correlations
between PNM and SNM, the uncertainty on $S_2(n)$ and $L(n)$ would be
overestimated.
More details can be found in Refs.~\cite{Melendez:2019izc,
Drischler:2020preparation}.

\begin{figure}[htb]
	\centering
 \includegraphics{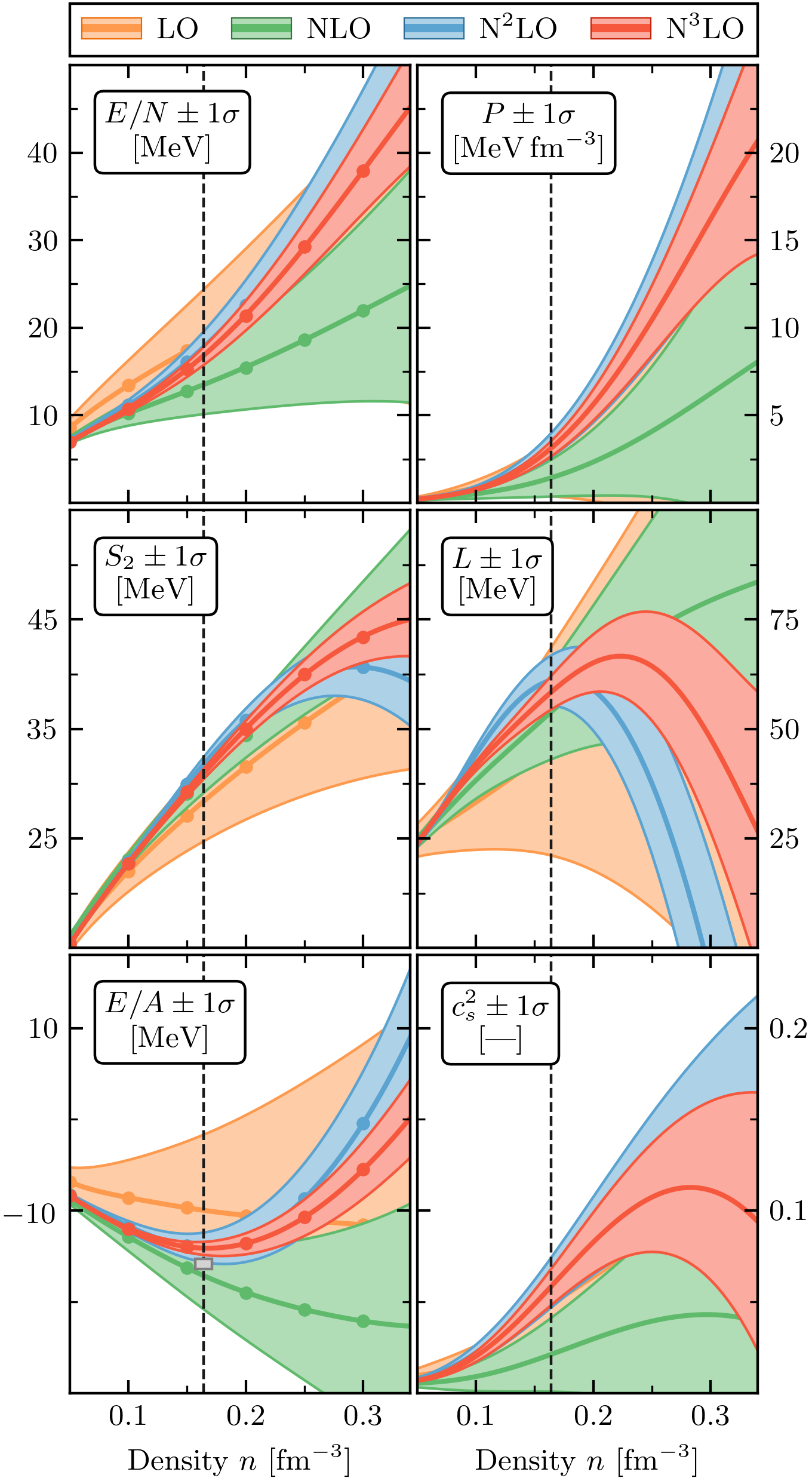}
    \phantomsublabel{-3.}{5.25}{fig:energy_per_neutron}
	\phantomsublabel{-1.54}{5.25}{fig:pressure}
	\phantomsublabel{-3.}{3.4}{fig:symmetry_energy}
	\phantomsublabel{-1.54}{3.4}{fig:slope}
	\phantomsublabel{-3.}{1.52}{fig:energy_per_particle}
	\phantomsublabel{-1.54}{1.52}{fig:speed_of_sound}
 \caption{
 Order-by-order predictions with 68\% bands
 for~\protect\subref{fig:energy_per_neutron} the energy per particle
 $\frac{E}{N}(n)$ and~\protect\subref{fig:pressure} the pressure $P(n)$ of
 PNM;~\protect\subref{fig:symmetry_energy} the symmetry energy $S_2(n)$
 and~\protect\subref{fig:slope} its (rescaled) density
 dependence~$L(n)$;~\protect\subref{fig:energy_per_particle} the energy per
 particle $\frac{E}{A}(n)$ of SNM; and~\protect\subref{fig:speed_of_sound} the
 speed of sound $c_s^2(n)$ of PNM, each as a function of density. Dots denote
 every fifth interpolation point, where $n = 0.05,\, 0.06,\, \dotsc,\,
 0.34\fmiq$. The gray box in~\protect\subref{fig:energy_per_particle} depicts
 the empirical saturation point, $n_0 = 0.164 \pm 0.007 \fmiq$ with
 $\frac{E}{A}(n_0) = -15.86 \pm 0.57 \MeV$, obtained from a set of energy
 density functionals~\cite{Drischler:2015eba,Drischler:2017wtt}. The vertical
 lines are located at $n = 0.164\fmiq$. See the main text for details.
 } \label{fig:energies_and_related_derivatives}
\end{figure}

\prlsection{Results}%
Figure~\ref{fig:energies_and_related_derivatives} shows our order-by-order
\chiEFT\ predictions, up to \NNNLO, for $\frac{E}{N}(n)$, $P(n)$, and $c_s^2(n)$
in PNM, as well as $S_2(n)$, $L(n)$, and $\frac{E}{A}(n)$.
We find an EFT breakdown scale $\Lambda_b$ consistent with $600\MeV$ and
optimized truncation-error correlation lengths $\ell = \verifyvalue{0.97}\fmi$
($\verifyvalue{0.48}\fmi$) for PNM (SNM). The correlation between the truncation
errors of $\frac{E}{N}(n)$ and $\frac{E}{A}(n)$ is $\rho =
\verifyvalue{0.94}$.
These hyperparameters are only tuned to the derivative-free quantities
$\frac{E}{N}(n)$ and $\frac{E}{A}(n)$; derivatives and their uncertainties are
thus \emph{pure predictions} of our framework, as are $S_2(n)$ and $L(n)$.
The bands in Fig.~\ref{fig:energies_and_related_derivatives} account for both
the EFT truncation error and the overall interpolation uncertainty. Our Bayesian
$1\sigma$ uncertainties for derivative-free quantities are broadly similar~\cite{Furnstahl:2015rha} to those obtained using the
``standard EFT'' error prescription~\cite{Phillips:1999hh,Griesshammer:2012we,Epel15NNn4lo, Epel15improved}, \eg, as
it was applied to UQ of the EOS in Ref.~\cite{Drischler:2017wtt}.  But only our
correlated approach can propagate these reliably to
$P(n)$, $S_2(n)$, $L(n)$,  and $c_s^2(n)$.

We observe an order-by-order EFT convergence pattern for the observables at low
densities, $n \lesssim 0.1\fmiq$. However, at \NNLO and beyond, 3N interactions
enter the \chiEFT\ expansion with repulsive contributions, especially at
densities $n \gg n_0$. Their \NNLO and \NNNLO EFT corrections then have a
markedly different density dependence, as indicated by our model-checking
diagnostics~\cite{Melendez:2019izc} for each energy per
particle~\cite{Drischler:2020preparation}.
This produces bands that do not appear to encapsulate higher-order predictions.
Nevertheless, we stress caution when critiquing the consistency of the
uncertainty bands; because of the strong correlations, statistical fluctuations can
occur over large ranges in density. Our credible interval diagnostics show that
the bands are consistent up to these
fluctuations~\cite{Drischler:2020preparation}.

The distributions of all observables follow a multivariate Gaussian, except for
$c_s^2(n)$, which requires sampling. The strong correlation between
$\frac{E}{N}(n)$ and $\frac{E}{A}(n)$ produces narrow constraints at $n_0$: $S_v
= \verifyvalue{31.7 \pm 1.1} \MeV$ and $L = \verifyvalue{59.8 \pm 4.1} \MeV$ at
the $1\sigma$ level. These agree remarkably well with central values from the analyses compiled in Ref.~\cite{Li:2019xxz}.
%
Our results for $c_s^2(n)$ are below the asymptotic high-density limit predicted
by perturbative quantum chromodynamics, $c_s^2\left(n \gg 50n_0\right) =
\frac{1}{3}$~\cite{Fraga:2013qra}. The uncertainties, however, are sizeable at
the maximum density: $c_s^2(2n_0)  \simeq \verifyvalue{0.14 \pm 0.08}$ (\NNLO)
and $c_s^2(2n_0) \simeq \verifyvalue{0.10 \pm 0.07}$ (\NNNLO). Precise
measurements of neutron stars with mass $\gtrsim 2\Msol$~\cite{Demo10ns,
Anto13PSRM201, Fons16pulsars, Cromartie:2019kug} indicate that the limit has to
be exceeded in some density regime beyond $n_0$~\cite{Bedaque:2014sqa}. Our
$2\sigma$ uncertainty bands are consistent with this happening slightly above
$2n_0$, especially since the downward turn of $c_s^2\left(n\gtrsim 0.28\fmiq\right)$
is likely an edge effect that will disappear if we train on data at even higher
densities.

\begin{figure}[htb]
 \centering
 \includegraphics{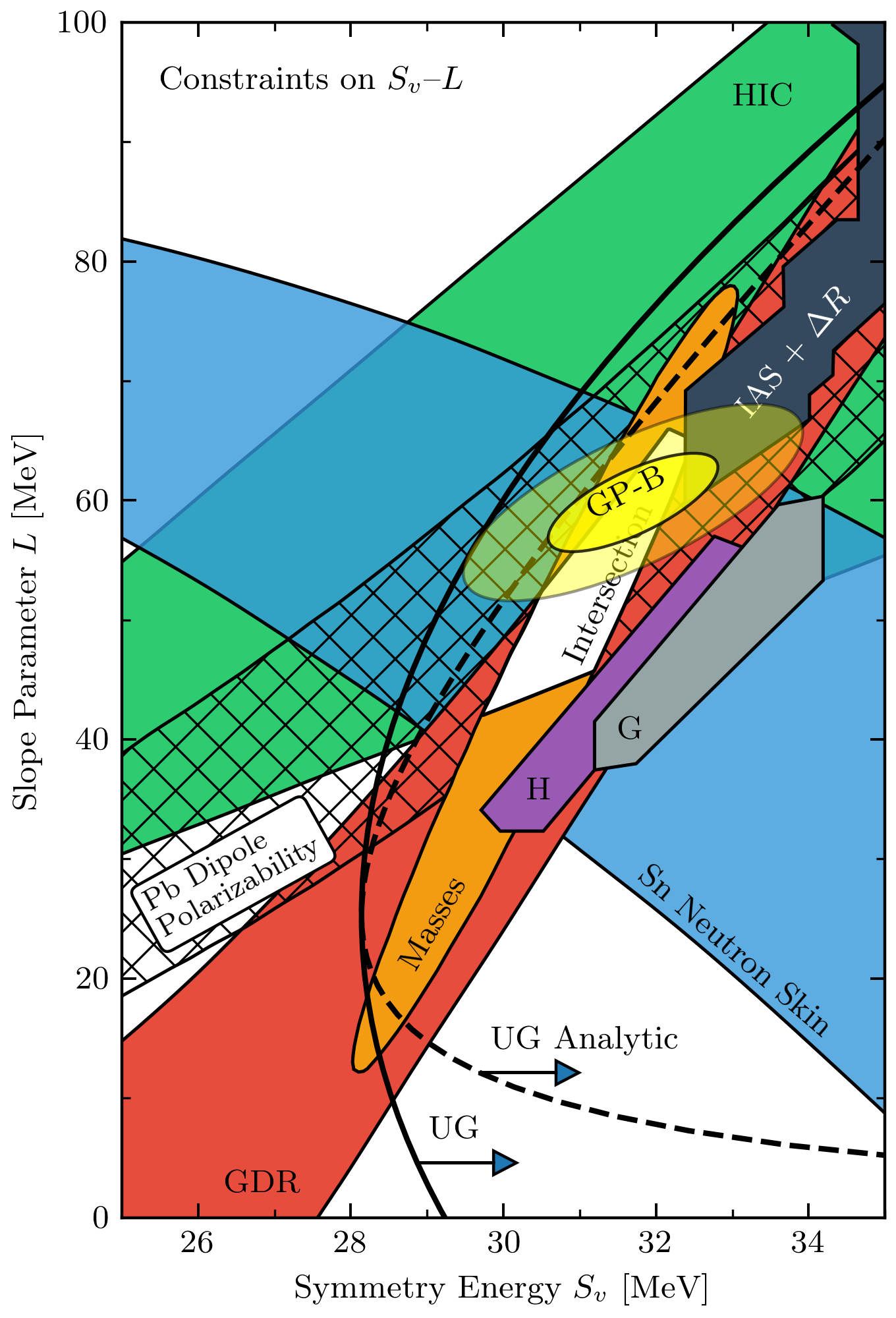}
 \caption{
 Constraints on the $S_v$--$L$ correlation. Our results (``GP--B'') are given at
 the 68\% (dark-yellow ellipse) and 95\% level (light-yellow ellipse).
 Experimental constraints are derived from heavy-ion collisions
 (HIC)~\cite{Tsan09Lconstr}, neutron-skin thicknesses of \isotope{Sn}
 isotopes~\cite{Chen10skin}, giant dipole resonances (GDR)~\cite{Trip08gdr}, the
 dipole polarizability of \isotope[208]{Pb}~\cite{Tami11dipole,Roca13dipoPb},
 and nuclear masses~\cite{Kort10edf}. The intersection is depicted by the white
 area, which only barely overlaps with constraints from isobaric analog states
 and isovector skins ($\text{IAS}+\Delta R$)~\cite{Dani17IVskins}. In addition,
 theoretical constraints derived from microscopic neutron-matter calculations by
 Hebeler~\etal~(H)~\cite{Hebe10PRL} and \mbox{Gandolfi}~\etal~(G)~\cite{Gand12nm} as
 well as from the unitary gas (UG) limit by Tews~\etal~\cite{Tews17NMatter}.
 The figure has
 been adapted from Refs.~\cite{Lattimer:2014sga,Lattimer:2012xj}. A Jupyter
 notebook that generates it is provided in Ref.~\cite{BUQEYEgithub}.
 } \label{fig:esym_L}
\end{figure}

\prlsection{Comparison with Experiment}%
Figure~\ref{fig:esym_L} depicts constraints in the $S_v$--$L$ plane. The allowed
region we derive from \chiEFT\ calculations of infinite matter is shown as the
yellow ellipses (dark: $1\sigma$, light: $2\sigma$) and denoted ``GP-B''
(Gaussian process--BUQEYE collaboration). Also shown are several experimental
and theoretical constraints compiled by
Lattimer~\etal~\cite{Lattimer:2014sga,Lattimer:2012xj,Tews17NMatter}. The experimental
constraints include measurements of isoscalar giant dipole resonances, dipole
polarizabilities, and neutron-skin thicknesses (see the caption for details).
The white area depicts the intersection of all these (excluding that from
isobaric analog states and isovector skins, which barely overlaps). This
region is in excellent agreement with our prediction.

Our yellow ellipses in Fig.~\ref{fig:esym_L} represent the posterior $\pr(S_v, L
\given \mathcal{D})$, where the training data $\mathcal{D}$ are the
order-by-order predictions of $\frac{E}{N}(n)$ and $\frac{E}{A}(n)$ up to
$2n_0$. The distribution is accurately approximated by a two-dimensional
Gaussian with mean and covariance
\begin{align} \label{eq:Sv_L_mean_cov}
\verifyvalue{
\begin{bmatrix}
\mu_{S_v} \\
\mu_L
\end{bmatrix}
=
\begin{bmatrix}
31.7 \\ 59.8
\end{bmatrix}
\quad \text{and} \quad
\Sigma =
\begin{bmatrix}
 1.11^2 & 3.27 \\
 3.27 & 4.12^2
\end{bmatrix}.
}
\end{align}
We consider all likely values of $n_0$ via $\pr(S_v, \,L \given \mathcal{D}) =
\int \pr(S_2, \,L \given n_0, \mathcal{D}) \pr(n_0 \given \mathcal{D})
\dd{n_0}$. Here, $\pr(S_2, \,L \given n_0, \mathcal{D})$ describes the
correlated to-all-orders predictions at a particular density $n_0$, and $\pr(n_0
\given \mathcal{D}) \approx \verifyvalue{0.17 \pm 0.01} \fmiq$ is the Gaussian
posterior for the saturation density, including truncation errors, determined in
Ref.~\cite{Drischler:2020preparation}. If the canonical empirical saturation
density, $n_0 = 0.164 \fmiq$, is used instead the posterior mean shifts
slightly downwards:
$S_v \to S_v - \verifyvalue{0.8}\MeV$ and $L \to L - \verifyvalue{1.4}\MeV$.
This shift is well within the uncertainties computed using our internally
consistent $n_0$. In contrast to experiments, which extract $S_v$--$L$ from
measurements over a range of densities, our theoretical approach predicts
directly at saturation density, thereby removing artifacts induced by
extrapolation.

Our $2\sigma$ ellipse falls completely within constraints derived from the
conjecture that the unitary gas is a lower limit on the EOS~\cite{Tews17NMatter}
(solid black line). The same work also made additional simplifying assumptions
to derive an analytic bound---only our $1\sigma$ ellipse is fully within that
region (dashed black line). Figure~\ref{fig:esym_L} also shows the allowed
regions obtained from microscopic neutron-matter calculations by
Hebeler~\etal~\cite{Hebe10PRL} (based on \chiEFT\ NN and 3N interactions fit to
few-body data only) and \mbox{Gandolfi}~\etal~\cite{Gand12nm} (where 3N interactions
were adjusted to a range of $S_v$). The predicted ranges in $S_v$ agree with
ours, but we find that $L$ is $\approx \verifyvalue{10}\MeV$ larger,
corresponding to a stronger density-dependence of $S_2(n_0)$.
References~\cite{Hebe10PRL,Gand12nm} quote relatively narrow ranges for
$S_v$--$L$, but those come from surveying available parameters in the
Hamiltonians and so---unlike our quoted intervals---do not have a statistical
interpretation.

\prlsection{Summary and Outlook}%
We presented a novel framework for EFT truncation errors that includes
correlations within and between observables. It enables the efficient
evaluation of derived quantities.
We then constrained multiple key observables for neutron-star physics based on
cutting-edge MBPT calculations with \chiEFT\ NN and 3N interactions up
to \NNNLO. Correlations in the EFT truncation error---both across densities and
between different observables---must be accounted for in order to obtain full
credible intervals. In several cases (\eg, $S_v$) the result is a much smaller
uncertainty than one might na\"ively expect. Our narrow predictions for
$S_v$--$L$ are in excellent agreement with the joint experimental constraint.

A rigorous comparison between empirical constraints on the EOS and our knowledge
of the underlying microscopic dynamics of strongly interacting nuclear matter is
particularly important in the era of multimessenger astronomy because new constraints on the neutron-star EOS are anticipated from NASA's
NICER~\cite{Bogd19NICER2, Riley19NICER, Mill19NICER, Raai19NICER} and from
the LIGO-Virgo collaboration~\cite{LIGO17MultMess, LIGO18NSradii, LIGO18update}. Our EOS results are in good agreement with recent observations, especially those with input from NICER (cf. theory-agnostic joint observational posteriors in Figure~1 of Ref.~\cite{Essick:2020flb}).
Nuclear physics experiments 
(\eg, those in Refs.~\cite{Horo14CREXPREX,Bala14FRIB,Birkhan:2016qkr,Kaufmann:2020gbf})
will also contribute important information to this
overall picture. 

Our Bayesian framework can be straightforwardly adapted and used in future studies that will more firmly establish
this comparison. A full Bayesian analysis can be performed via Markov Chain
Monte Carlo sampling over GP hyperparameters and the low-energy couplings in the
nuclear interactions~\cite{Carlsson:2015vda, Wesolowski:2015fqa,
Wesolowski:2018lzj}. This requires the development of improved chiral NN and
3N forces up to \NNNLO~\cite{Hoppe:2019uyw,Huth19int,Epelbaum:2019kcf}. Extensions to arbitrary
isospin asymmetry and finite temperature are also an important avenue for future
study. Work in all these directions will be facilitated by the public
availability of the tools presented here as Jupyter
notebooks~\cite{BUQEYEgithub}.

\begin{acknowledgments}

We thank S.~Reddy for fruitful discussions. We are also grateful to
the organizers of ``Bayesian Inference in Subatomic Physics---A Marcus Wallenberg 
Symposium'' at Chalmers University of Technology, Gothenburg for creating a 
stimulating environment to learn and discuss the use of statistical methods in 
nuclear physics. C.D. acknowledges support by the
Alexander von Humboldt Foundation through a Feodor-Lynen Fellowship and the U.S.
Department of Energy, the Office of Science, the Office of Nuclear Physics, and
SciDAC under Awards DE-SC00046548 and DE-AC02-05CH11231. The work of R.J.F. and J.A.M.
was supported in part by the National Science Foundation under Grant
Nos.~PHY--1614460 and PHY--1913069, and the NUCLEI SciDAC Collaboration under U.S.
Department of Energy MSU subcontract RC107839-OSU\@. The work of D.R.P was
supported by the U.S. Department of Energy under Award DE-FG02-93ER-40756 and
by the National Science Foundation under PHY-1630782, N3AS FRHTP\@. C.D. thanks
the Physics Departments of The Ohio State University and Ohio University for
their warm hospitality during extended stays in the BUQEYE state.

\end{acknowledgments}

\bibliographystyle{apsrev4-1}
\bibliography{bib/bayesian_refs,bib/additional}

\end{document}